\documentclass[a4paper]{jpconf}
\usepackage{graphicx}

\begin{document}

\title{Mott glass phase in a diluted bilayer Heisenberg quantum antiferromagnet}

\author{Nv-Sen Ma$^1$, Anders W. Sandvik$^2$, and Dao-Xin Yao$^1$}
\address{$^1$State Key Laboratory of Optoelectronic Materials and
Technologies, School of Physics and Engineering, Sun Yat-sen
University, Guangzhou 510275, People's Republic of China}
\address{$^2$Department of Physics, Boston University, 590 Commonwealth Avenue, Boston, Massachusetts 02215, USA}
\ead{sandvik@bu.edu,~~yaodaox@mail.sysu.edu.cn}

\begin{abstract}
We use quantum Monte Carlo simulations to study a dimer-diluted $S=1/2$ Heisenberg model on a bilayer square lattice with intralayer interaction
$J_{1}$ and interlayer interaction $J_{2}$. Below the classical percolation threshold $p_c$,
the system has three phases reachable by tuning the interaction ratio $g=J_{2}/J_{1}$: a N$\acute{e}$el ordered phase, a gapless quantum
glass phase, and a gapped quantum paramagnetic phase. We present the ground-state phase diagram in the plane of dilution $p$ and interaction
ratio $g$. The quantum glass phase is certified to be of the gapless Mott glass type, having a uniform susceptibility vanishing at zero
temperature $T$ and following a stretched exponential form at $T>0$; $\chi_u \sim \exp(-b/T^{\alpha})$ with $\alpha < 1$. At the phase transition
point from N$\acute{e}$el ordered to Mott glass, we find that the critical exponents are different from those of the clean system
described by the standard $O(3)$ universality class in 2+1 dimensions.
\end{abstract}

\section{Introduction}
\vskip2mm

\begin{figure}[h]
\begin{center}
\includegraphics[width=18pc]{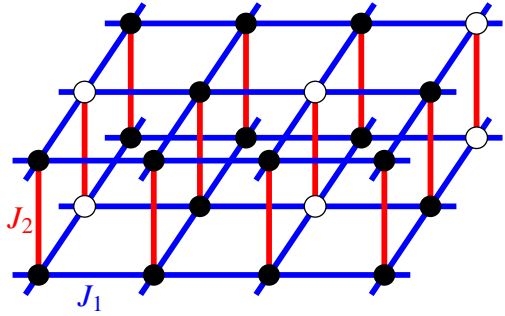}\hspace{2pc}
\end{center}
\vskip-5mm
\caption{The bilayer Heisenberg model with intraplane coupling $J_{1}$ (blue bonds) and interplane coupling $J_{2}$ (red vertical bonds).
The open circles stand for removed dimers (i.e., the two spins are removed, not just the coupling between them). By tuning the coupling ratio
$g=J_{2}/J_{1}$ a QPT can be reached for any dilution fraction $p<p_c$, with $p_c \approx 0.407$ \cite{percocri} being the classical percolation point.}
\label{model}
\end{figure}

The effects of disorder (randomness) on the critical behavior represent an important aspect of quantum phase transitions (QPTs). Disorder brings
with it a plethora of profound phenomena beyond those present in clean systems~\cite{cusuper,fermi}, with many open questions of broad interest
in quantum many-body physics. Many different kinds of disorders can be considered, e.g., random dilution of the degrees of freedom or random
variations in the coupling constants. Among the possible disorder channels in quantum spin systems, random dilution has attracted special attention
because of the possibilities to achieve it experimentally by substitution of magnetic ions by non-magnetic ones, e.g., replacing Cu bu Zn or
Mg in the high-T$_{\rm c}$ cuprates \cite{experiment}. The case is also of particular interest from a theoretical point of view as there is simple notion
of classical ``geometrical'' fluctuations which interplay with the quantum fluctuations ~\cite{multicri}. Thus, a multicritical point can appear
if those two fluctuations diverge together, as a classical percolation point is approached while at the same time tuning the quantum fluctuations.
A prototypical example is the relatively well understood transverse-field Ising model with dimensionality $d>1$~\cite{multicri,transising}.  Many
works have also shown that a multicritical point can be realized in the diluted bilayer Heisenberg model with different intralayer interaction
$J_{1}$ and interlayer interaction $J_{2}$, if the dilution is accomplished by random removal of entire dimers (instead of diluting individual
spins, which leaves behind ``dangling'' spins which lead to different physics)~\cite{dilution1,dilution2}. The bilayer model is illustrated in
Fig.~\ref{model} and is further studied in this paper.

\begin{figure}[h]
\begin{center}
\includegraphics[width=16pc]{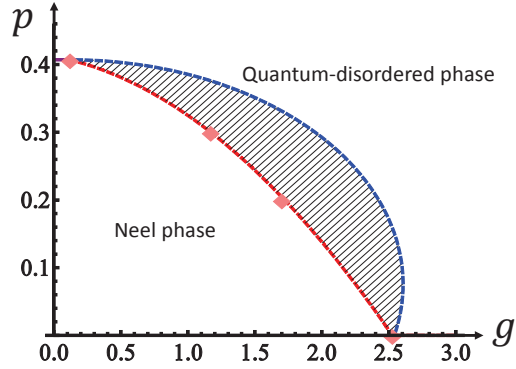}
\end{center}
\vskip-5mm
\caption{Phase diagram of the diluted bilayer Heisenberg model. The shaded area stands for the MG phase existing between the N$\acute{e}$el
and quantum-disordered phases.}
\label{phasedia}
\end{figure}

According to previous studies, at the classical percolation point, $p_c \approx 0.407$, the percolating cluster remains ordered until the
quantum critical point located at $g_{c}^{*}=0.118(6)$ ~\cite{dynamicalz,dilution1}. 
In this paper, we study the QPT when $p<p_{c}$ and obtain the phase diagram
at $T=0$ (shown in Fig.~\ref{phasedia}) using the efficient stochastic series expansion (SSE) quantum Monte Carlo (QMC) method \cite{sse}. 
While this criticality has been studied previously \cite {dilution1,dilution2}, we here re-examine it in light of recent studies 
\cite{harris2,j1j2j3} of other, similar quantum phase transitions where no changes in the critical exponents were found in the presence of 
disorder---in violation of the Harris criterion \cite{harris} for the relevance of disorder. In addition,
we study ``quantum glass'' (Griffiths) behavior away from the critical curve. Such phases come in two main variants in random quantum spin
systems and boson systems: the Bose glass (BG) and Mott glass (MG) \cite{mgbg}. Both of them are gapless, with the BG having a non-zero uniform
susceptibility $\chi_u$ (compressibility in Boson language, where the magnetization of a spin system is mapped onto the density of the
Boson system) at $T=0$ while the MG has a vanishing $\chi_u$ at $T=0$ \cite{diffglass}. It is believed that the MG can only exist in random systems
with particle-hole symmetry \cite{bringin}, though recently evidence has been found of similar behavior also without this symmetry (an MG or
and effective MG with extremely small essentially undetectable $T=0$ compressibility) \cite{wang15}. In our previous work, the MG phase was investigated in a
two-dimensional square-lattice dimerized Heisenberg model with three different nearest-neighbor couplings \cite{j1j2j3} and the suscetibility was
found to follow a stretched exponential form. We here find the same characteristic form in the diluted bilayer.

\section{Model and critical behavior}
\vskip2mm

The Hamiltonian of the $S=1/2$ spin model illustrated in Fig.~\ref{model} is give by
\begin{equation}
H=J_{1}\sum_{a=1,2}\sum_{\langle ij\rangle} {\bf S}_{i,a}\cdot {\bf S}_{j,a}+J_{2}\sum_{i} {\bf S}_{i,1}\cdot {\bf S}_{i,2},
\label{hamiltonian}
\end{equation}
where $a=1$ $(2)$ stands for the upper (lower) plane of the bilayer system. Dimers of spins, one in the upper layer and one in the lower layer
(coupled by the intra-dimer coupling $J_2$), are removed at random with probability $p <p_c$. We normalize by setting the intra-plane interaction
$J_{1}=1$. In the clean system, $p=0$, the system goes through a QPT between N$\acute{e}$el to quantum disordered phase at the critical point
located at $g_{c}(p=0)=2.525(2)$ ~\cite{cleangc}. The critical exponents are the same as the classical $3D$ Heisenberg model, i.e., the 3D O(3) universality
class.  When $g=0$, the model turns into two separate diluted $2D$ square lattice Heisenberg and at the classical percolation point
$p_{c} \approx 0.407$
the percolating cluster remains ordered at $T=0$. This order persists in the coupled layers as well up to $g_c(p=p_c) \approx 0.11$, which is
a multi-critical point with simultaneous critical geometrical (percolation) fluctuations and critical quantum fluctuations of the spins
on those classically critical clusters.

We here consider the system between the special points $g_c(p=0)$ and $g_c(p=p_c)$, on the critical curve as well as for larger $g$, in the shaded
glass region in the phase diagram of Fig.~\ref{phasedia}. This is a preliminary report of results of SSE QMC calculations on lattices as large as $N=2\times64\times64$
sites, averaged over $100$ dilution configurations in each case. For a given disorder realization we identify the largest cluster and only use it in the
simulations. We take advantage of the ``$\beta$-doubling'' technique \cite{bdouble} 
to quickly equilibrate the system to the very low temperatures needed to study the ground state.

\begin{figure}[h]
\begin{center}
\includegraphics[width=30pc]{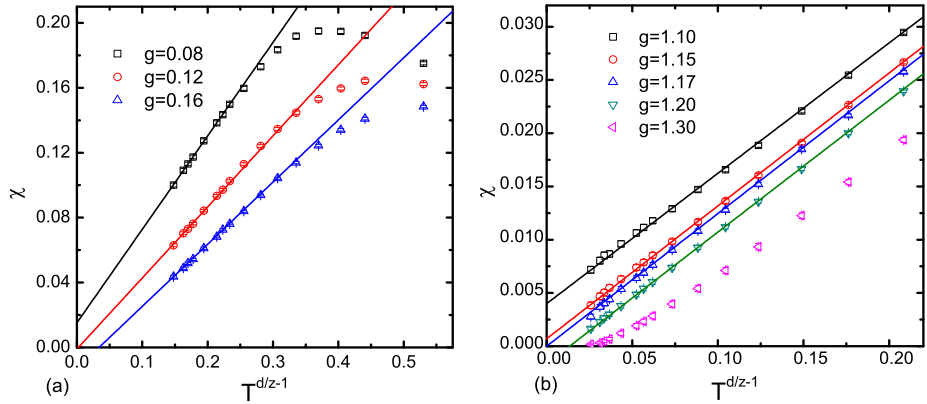}
\end{center}
\vskip-5mm
\caption{Analysis of the uniform susceptibility delivering the dynamic exponent by adjusting the value of $z$ for the best linear behavior
of $\chi_u$ versus $T^{2/z-1}$ for coupling ratios $g$ close to the QPT. In (a) $p=p_{c}=0.407$ while in (b) $p=0.3$. At $p_{c}$ the fitted line passes
through $(0,0)$ when $g \approx 0.12$ and $z \approx 1.36$. For $p=0.3$ we observe linear behavior when $z \approx 1.14$ and locating the zero intercept
(indicating the QPT) gives $g_{c}\approx1.17$.}
\label{critical}
\end{figure}

\section{Dynamic exponent}
\vskip2mm

Close to a quantum-critical point, the uniform susceptibility in the thermodynamic limit should have the following characteristic form \cite{jingwu}:
\begin{equation}
 \chi_{u}=a+b/T^{d/z-1},
\label{relationx}
\end{equation}
where $a$ and $b$ are constants, with $a=0$ at $g_{c}$, and $d$ is the dimensionality which at $p_c$ in the cluster fractal dimensionality,
$d=91/48$ \cite{percocri}, and for $p<p_c$ $d=2$. We calculate $\chi_{u}$ in the simulations using the standard expression
\begin{equation}
 \chi_{u}=\frac{\beta}{N}\left \langle \left (\sum_{i=1}^{N}S_{i}^{z} \right )^{2}\right \rangle.
\label{uniformx}
\end{equation}
To test the form (\ref{relationx}) we follow the $T$-dependence for different system sizes and only report results down to the $T$ for which good 
convergence with $L$ has been achieved.

From Eq.~(\ref{relationx}), we can see that $\chi$ has a linear relation with $T^{d/z-1}$. If $z$ is not known, we can use it as an adjustable parameter
and optimize it for linear scaling. In addition, when a line fitted to the data has vanishing intercept $a$ the $g$ value will be the critical one.
Results close to the multicritical point ($p_{c}$,$g_{c}$) are shown in Fig.~\ref{critical}(a). We here use $z=1.36$ as obtained in Ref.~\cite{dynamicalz}
and see that the critical $g_{c} \approx 0.12$, which is consistent with the previous results. This procedure has not previously been considered for
$p<p_c$ and we show results at $p=0.3$ in Fig.~\ref{critical}(b), where we use $d=2$ and adjust $z$ to achieve a linear behavior. We again observe remarkably
good scaling for several $g$ values, but now with $z=1.14(2)$. The critical point is at $g_{c}\approx1.17$ for $p=0.3$. Note that a cross-over away from
the linear behavior can be seen for $g>g_c(p)$ but for $g<g_c$ we do not observe any cross-overs at the temperatures studied. At $p_c$ it has been
speculated that the scaling may hold all the way down to $T=0$ because the percolating cluster has vanishing stiffness (which sets the scale
of the cross-over) \cite{dynamicalz}, but for $p<p_c$ there must eventually be a cross-over because the stiffness is non-zero.

We have also studied the critical behavior at $p=0.2$ and find the critical point $g_{c}\approx1.70$ with $z=1.08(3)$. Considering the error bars,
the $z$-value is not very different from the value obtained at $p=0.3$, and the most likely scenario is that $z$ is the same but different from $1$ for
all $p < p_c$, which would agree with Ref.~\cite{dilution2}, where the critical exponents were found not not change with $p$ in the classical model.
In that study, a scaling correction was used because the fitting curve deviates significantly from the pure power-law behavior. However, in our
case the power-law fitting curves are of high quality with $\chi^{2}\approx 1$ for $g=1.17$ in Fig.~\ref{critical}. The dynamic exponent is also
quite different from the random-coupling square lattice Heisenberg model, where $z=1$ always appears to hold \cite{j1j2j3}. This implies that the two
kinds of disordered systems have quite different low-energy physics systems, which would not be expected at first sight and for which there is
no theoretical explanation as of yet.

The Harris criterion can normally be used to determine the relevance of disorder at classical phase transitions. It says that the correlation length
exponent $\nu$ in the clean system has to fulfill $\nu\geq 2/d$ if disorder is irrelevant. Otherwise, the critical exponents will change when disorder is
introduced in order to satisfy the inequality \cite{harris}. Although it has been thought to be applicable to QPTs, several disordered systems have
shown violations of the Harris criterion.~\cite{j1j2j3,harris2} The dimensionality $d$ for a QPT in Harris criterion is the number of dimensions
which have disorder. Thus, in this model $d=2$ and disorder is expected to be relevant to the clean bilayer antiferromagnet which has $\nu\approx0.705<1$,
as we have found here (unlike the system in Ref.~\cite{j1j2j3}). Although we have not yet computed $\nu$, we note that the change in $z$ indicates a
new universality class.

\section{Mott-Glass Phase}
\vskip2mm

\begin{figure}[h]
\begin{center}
\includegraphics[width=20pc]{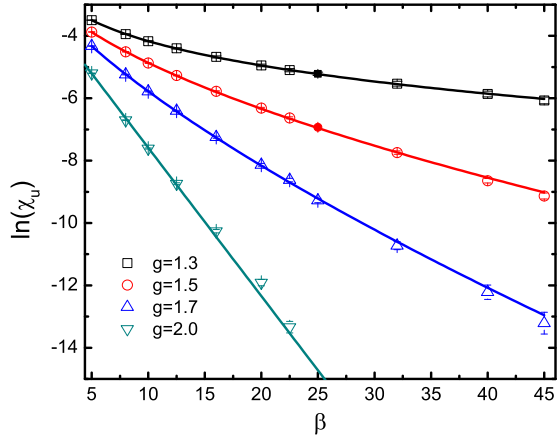}
\end{center}
\vskip-5mm
\caption{Inverse-temperature dependence of the logarithm of the magnetic susceptibility at $p=0.3$ and $g$ larger than the critical value.
When $g=1.3, 1.5$ and $1.7$, the behaviors (fitted curves) follow the form (\ref{glassx}) with $\alpha=0.21(1)$, $0.58(1)$, and $0.71(2)$,
respectively. When $g=2$, the linear fitting illustrates the standard decay of the form (\ref{gapx}) with a gap $\Delta$.}
\label{glass}
\end{figure}

When disorder is introduced in, a new phase, often called a ``quantum glass'' should emerge. The uniform susceptibility $\chi_{u}$ should approach 
a nonzero value at low $T$ in N$\acute{e}$el phase, while in the gapped quantum-disordered phase
\begin{equation}
  \chi_{u}\sim {\rm exp}\left (-\frac{\Delta}{T} \right ),
\label{gapx}
\end{equation}
with the spin gap $\Delta$. If the system is in the gapless quantum glass phase of the Mott type, $\chi_u$ vanishes at zero temperature, and
the following form has been found empirically:
\begin{equation}
 \chi_{u}\sim {\rm exp}\left (-\frac{b}{T^{\alpha}} \right),
\label{glassx}
\end{equation}
where $\alpha=1/2$ in a diluted $S=1$ system~\cite{newmg,sone} and $0<\alpha<1$ in a disordered $S=1/2$ Heisenberg square lattice.~\cite{j1j2j3}
Accordingly, the temperature dependence  of $\ln(\chi_u)$ is a useful quantity to distinguish different ground states and test the existence of the MG state.
The results for $p=0.3$ when $g>g_{c}$ are presented in Fig.~\ref{glass}. It can be observed that $\chi_{u}$ follows the form (\ref{glassx}) with $0<\alpha<1$
when $g>g_{c}$ but not far away from the critical point. This implies that the MG phase exists in this bond diluted bilayer Heisenberg antiferromagnet.
The form (\ref{glassx}) also holds $p=0.2$.

\section{Summary}
\vskip2mm

We have studied the critical behavior of a dimer-diluted bilayer $S=1/2$ 
Heisenberg model below the classical percolation point using the SSE Monte Carlo simulation
method. Based on our calculations, we obtain the phase diagram shown in Fig.~\ref{phasedia}. For $p=0.3$, the QPT at zero temperature occurs at $g\approx1.17$
and the uniform susceptibility changes with temperature by following the power law form $\chi=a+b/T^{d/z-1}$, with $d=2$. Below the percolation threshold,
the dynamic exponent $z=1.18(3)$ was found. It is different from the clean case, where $z=1$, and consistent with the Harris criterion, according to which
the disorder of bond dilution is relevant to the system and the new critical exponents different from the clean case could appear. We have also calculated
$z$ for other disorder strengths and found no significant changes with $p$, though the claculation becomes difficult for $p<2$ due to slow approach
(with the system size) to the asymptotic critical scaling behavior. From the analysis of critical points, we find the gapless MG phase with vanishing
$\chi_{u}$ above the critical $g_{c}$ for $p<p_{c}$, where a stretched exponential relationship of $\chi$ versus $T$ holds.

\ack
We would like to thank Wenan Guo for stimulating discussions.
NSM and DXY acknowledge support from National Basic Research Program of China (2012CB821400), NSFC-11275279, RFDPHE of China (20110171110026), Fundamental Research Funds for the Central Universities of China, and NCET-11-0547. The work of AWS was supported by the NSF under Grants No.~DMR-1104708 and DMR-1410126.

\end{document}